\documentclass[10pt,conference]{IEEEtran}

\usepackage{graphicx}
\usepackage{dcolumn}
\usepackage{bm}
\usepackage{xcolor}
\usepackage{qcircuit}
\usepackage{braket}
\usepackage{float}
\usepackage{algorithm}
\usepackage{algorithmicx}
\usepackage{algpseudocode}
\usepackage{mathtools}
\usepackage{dirtytalk}
\usepackage{hyperref}
\usepackage{amsmath}
\usepackage{amsfonts}
\usepackage{amssymb}
\usepackage{amsthm}
\usepackage{newlfont}
\usepackage{array}
\usepackage{amsmath,amssymb,dsfont}

\usepackage{cite}
\usepackage{dsfont}
\usepackage{booktabs} 
\usepackage{cite}
\usepackage{subcaption}
\usepackage{pgfplots}
\usepackage{pgfgantt}
\definecolor{bluegantt}{HTML}{74BBC9}
\definecolor{yellowgantt}{HTML}{F7E967}

\pgfplotsset{compat=1.18}
\usepackage{tikz}
\usepackage{xcolor}
\definecolor{lgreen}{RGB}{80,200,120}   
\definecolor{lred}{RGB}{255,102,102}    
\definecolor{nred}{RGB}{249,115,118}    
\definecolor{mintgreen}{RGB}{143,213,191}
\definecolor{nviolet}{RGB}{182,142,242}
\definecolor{lblue}{RGB}{102,153,255}   
\definecolor{lyellow}{RGB}{255,204,102} 
\definecolor{lighteryellow}{RGB}{255,230,153} 
\definecolor{g1}{HTML}{7FC97F}
\definecolor{g2}{HTML}{FDC086}
\definecolor{g3}{HTML}{BEAED4}
\definecolor{g4}{HTML}{FFFF99}
\definecolor{g5}{HTML}{386CB0}
\definecolor{g6}{HTML}{F0027F}
\usepackage{amsmath}
\definecolor{forceblue}{RGB}{36,54,66} 
\usepackage[T1]{fontenc}

\usepackage{setspace}

\algnewcommand{\algorithmicand}{\textbf{ and }}
\algnewcommand{\algorithmicor}{\textbf{ or }}
\algnewcommand{\OR}{\algorithmicor}
\algnewcommand{\AND}{\algorithmicand}

\usepackage{authblk}

\title{Deadline-Aware Scheduling of Distributed Quantum Circuits in Near-Term Quantum Cloud}
\author[1]{Nour Dehaini}
\author[1]{Christia Chahoud}
\author[1]{Mahdi Chehimi}
\affil[1]{American University of Beirut}

\affil[ ]{\textit{nad29@mail.aub.edu, cec13@mail.aub.edu, mc127@aub.edu.lb}}

\begin{document}

\newcommand{\SP}[1]{\textcolor{blue}{Sharooz: #1}}
\newcommand{\MC}[1]{\textcolor{purple}{Mahdi: #1}}
\newcommand{\NP}[1]{\textcolor{green}{Nitish: #1}}
\newcommand{\DT}[1]{\textcolor{blue}{Don: #1}}

\maketitle
\begin{abstract}
Distributed quantum computing (DQC) enables scalable quantum computations by distributing large quantum circuits on multiple quantum processing units (QPUs) in the quantum cloud. In DQC, after partitioning quantum circuits, they must be scheduled and executed on heterogenous QPUs while balancing latency, overhead, QPU communication resource limits. However, since fully functioning quantum communication networks have not been realized yet, near-term quantum clouds will only rely on local operations and classical communication (LOCC) settings between QPUs, without entangled quantum links. Additionally, existing DQC scheduling frameworks do not account for user-defined execution deadlines and adopt inefficient wire cutting techniques. Accordingly, in this work, a \emph{deadline-aware DQC scheduling framework} with efficient wire cutting for near-term quantum cloud is proposed. The proposed framework schedules partitioned quantum subcircuits while accounting for circuit deadlines and QPU capacity limits. It also captures dependencies between partitioned subcircuits and distributes the execution of the sampling shots on different QPUs to have efficient wire cutting and faster execution. In this regard, a deadline-aware circuit scheduling optimization problem is formulated, and solved using simulated annealing. Simulation results show a marked improvement over existing \emph{shot-agnostic} frameworks under urgent deadlines, reaching a 12.8\% increase in requests served before their deadlines. Additionally, the proposed framework serves 8.16\% more requests, on average, compared to state-of-the-art \emph{dependency-agnostic} baseline frameworks, and by 9.60\% versus the \emph{dependency-and-shot-agnostic} baseline, all while achieving a smaller makespan of the DQC execution. Moreover, the proposed framework serves 23.7\%, 24.5\%, and 25.38\% more requests compared to \emph{greedy}, \emph{list scheduling}, and \emph{random} schedulers, respectively.
\end{abstract}

\IEEEpeerreviewmaketitle

 \vspace{-0.2cm}
\section{Introduction}\vspace{-0.1cm}

Quantum cloud services incorporate several quantum processing units (QPUs) dedicated to serve user-submitted requests to execute quantum circuits for different quantum applications. However, today’s noisy, intermediate-scale quantum (NISQ) QPUs have a limited number of noisy qubits and cannot execute deep or large-scale quantum circuits \cite{BARRAL2025100747}. In this regard, distributed quantum computing (DQC) is a framework that overcomes these limitations by partitioning more complex quantum circuits and executing them on multiple QPUs \cite{caleffi2024distributed}. Thus, DQC allows the execution of larger quantum circuits than what any single QPU device can execute alone.  

This interconnection of QPUs requires the quantum cloud to have quantum communication networks to exchange quantum states between the different devices through entanglement distribution, teleportation, and related protocols \cite{caleffi2024distributed}. However, these networking capabilities are still at an early stage, limiting the near-term scalability of DQC in cloud-based settings \cite{chehimi2022physics}. As a result, recent works have shifted toward exploring \emph{circuit cutting} and \emph{local operations and classical communication} (LOCC) as near-term quantum cloud DQC enablers without requiring quantum communication links between QPUs \cite{BARRAL2025100747}.

In this circuit cutting compiling step of the quantum cloud, large quantum circuits are partitioned into smaller subcircuits that can be executed independently and \emph{classically} recombined \cite{Peng_2020,gate_cut}. The two most commonly considered techniques are \emph{wire cutting} \cite{Peng_2020}, which cuts qubit wires and replace them with independent local state preparations and measurements, and \emph{gate cutting} \cite{gate_cut}, which decomposes entangling gates into local operations with probabilistic reconstruction \cite{HPC}. Both methods enable distributed execution using local operations, though at the cost of sampling overhead that scales exponentially with the number of cuts. The partitioned subcircuits must then be scheduled and executed on the cloud's QPUs while accounting for the user-specific requirements and QPU constraints.

Additionally, to reduce the number of generated subcircuits and reduce the sampling overhead, an LOCC wire-cutting technique was recently developed \cite{LOCC_wirecut}, which requires classical coordination between measurement and preparation subcircuits. However, this technique introduces dependencies between the subcircuits, which makes the task of DQC scheduling more challenging because of the several resulting constraints. When multiple users submit large quantum circuits to be partitioned, distributed, and scheduled across heterogeneous QPUs, the scheduler must balance competing objectives: minimizing execution time, respecting subcircuits dependencies and device constraints, and meeting user-specific deadlines, which makes the design and optimization of the quantum cloud scheduler a challenging task.

\subsection{Related Works}

Prior works have examined some of the aforementioned DQC scheduling challenges in isolation, but, to the best of our knowledge, no unified framework was developed to jointly address them in near-term quantum cloud. In general, recent efforts have explored DQC scheduling in quantum cloud settings with quantum communication networks between QPUs \cite{10821222,ferrari2024design,bahrani2024resource,chandra2024network,EC2S,larqucut}. For instance, the work in \cite{10821222} investigated how to schedule DQC jobs across QPUs utlizing entanglement distribution, while the work in 
\cite{ferrari2024design} presented a simulation framework for DQC that integrates computational and quantum networking aspects. Additionally, the work in \cite{bahrani2024resource} focused on resource management and circuit scheduling for data-center–scale heterogeneous QPU quantum communication networks. Moreover, in \cite{chandra2024network}, the authors proposed scheduling frameworks for quantum network operations required to support non-local gates in DQC. Furthermore, the work in \cite {EC2S} proposed a framework implementing circuit cutting techniques with entanglement-based distribution on QPUs, and the work in \cite{larqucut} proposed a circuit cutting and mapping framework for large-size circuits, using wire cuts and gate cuts with remote entanglement between nodes to avoid fully independent subcircuits which minimizes the sampling overhead. However, none of the works in \cite{10821222, ferrari2024design, bahrani2024resource,chandra2024network,EC2S,larqucut} considered near-term quantum cloud with LOCC only, as they all assumed the presence of shared entanglement between QPUs. Additionally, none of these works considered user-specified deadlines for the submitted quantum circuits, which is critical to represent real quantum applications. On top of that, all these works consider the execution of all sampling shots of the partitioned subcircuits on the same QPU, which is not optimized and can result in inefficiencies in DQC execution.

For near-term LOCC-based quantum cloud, recent works considered LOCC circuit cutting to solve the challenges of automating optimal cut placements and reducing sampling overhead \cite{optimal_wire&gate,LOCC_wirecut,gate_cut}, but without considering the resulting circuit scheduling problems. Additionally, some recent efforts explored DQC scheduling in these near-term cloud setups \cite{luo2025adaptive,Notads,HPC,cutandshoot}. For instance, the authors in \cite{luo2025adaptive} investigated adaptive job scheduling for DQC in cloud environments on a set of QPUs connected via LOCC and compared between four scheduling techniques. In \cite{Notads}, the authors introduced an integer linear programming scheduler in an LOCC-only setting that maps the subcircuits across QPUs, while respecting device-specific runtime limits. Similarly, in \cite{HPC}, the authors provide the Qdislib library for circuit cutting and focus on the parallel execution of quantum subcircuits across CPUs, GPUs, and QPUs via PyCOMPS framework. Nevertheless, none of the works in \cite{luo2025adaptive,Notads,HPC} considered user-definied deadlines nor developed a deadline-aware DQC job scheduling framework. Additionally, the works in \cite{luo2025adaptive,Notads,HPC} consider the execution of all sampling shots of the partitioned subcircuits on the the same QPU, which is not optimized and could result in delays in the DQC job execution time.  



In this regard, the work in \cite{cutandshoot} is the only existing work to address distributing both partitioned circuits and their sampling shots across multiple QPUs, analyzing its effect on minimizing the runtime. However, the shot distribution framework in \cite{cutandshoot} is not optimized and only allocates the shots uniformly over available QPUs or via user-specified rules, and, again, user-defined deadlines were not considered in \cite{cutandshoot}.

To the best of our knowledge, \emph{no existing work considered DQC scheduling in near-term multi-user quantum clouds based on LOCC only, where users specify the required application-specific execution deadlines for their submitted quantum circuits. Furthermore, no existing framework considered optimizing the subcircuit sampling shot distribution across heterogeneous QPUs to minimize execution time.}

\subsection{Contributions}
The main contribution of this work is the first deadline-aware DQC scheduling framework that supports multi-user workloads in a near-term quantum cloud with heterogeneous QPUs under an LOCC-only setting. In particular, we propose the first model that jointly considers circuit deadlines, efficient wire cutting, and sampling shots distribution in DQC scheduling. Towards this goal, we make key contributions:
\begin{itemize}
    \item We introduce a novel DQC scheduling framework that, unlike state-of-the-art works, adopts the recently-developed shot efficient LOCC wire-cut technique \cite{LOCC_wirecut} in an LOCC-based near-term quantum cloud. Particularly, we capture dependencies between the measurement and prepare subcircuits and develop the first framework to optimize the subcircuits' scheduling under these realistic constraints.   
    \item We formulate a deadline-aware DQC scheduling optimization framework that allows the distribution and optimization of subcircuits' sampling shots across heterogeneous QPUs, with different maximum possible circuit depths and qubits, while accounting for LOCC wire-cut dependencies and QPU eligibility and capacity limits, with the objective of maximizing the number of served user requests that meet their deadlines.
    \item We conduct extensive simulations comparing our \emph{proposed} framework with \emph{greedy}, \emph{list}, and \emph{random} schedulers, as well as state-of-the-art \emph{shot-agnostic} and \emph{dependency-agnostic} baselines. Under urgent deadlines, the \emph{proposed} framework serves 12.8\% more requests than \emph{shot-agnostic} scheme, and with mixed deadlines, it serves 8.16\% and 9.60\% more requests than dependency-agnostic and dependency-and-shot-agnostic baselines, while also reducing makespan, and 23.7\%, 24.5\%, and 25.38\% more requests than \emph{greedy}, \emph{list}, and \emph{random} schedulers, respectively.
\end{itemize}




\begin{figure*}[!t]
\begin{center}
\centering 
\hspace*{0.6cm}\includegraphics[width=\textwidth]{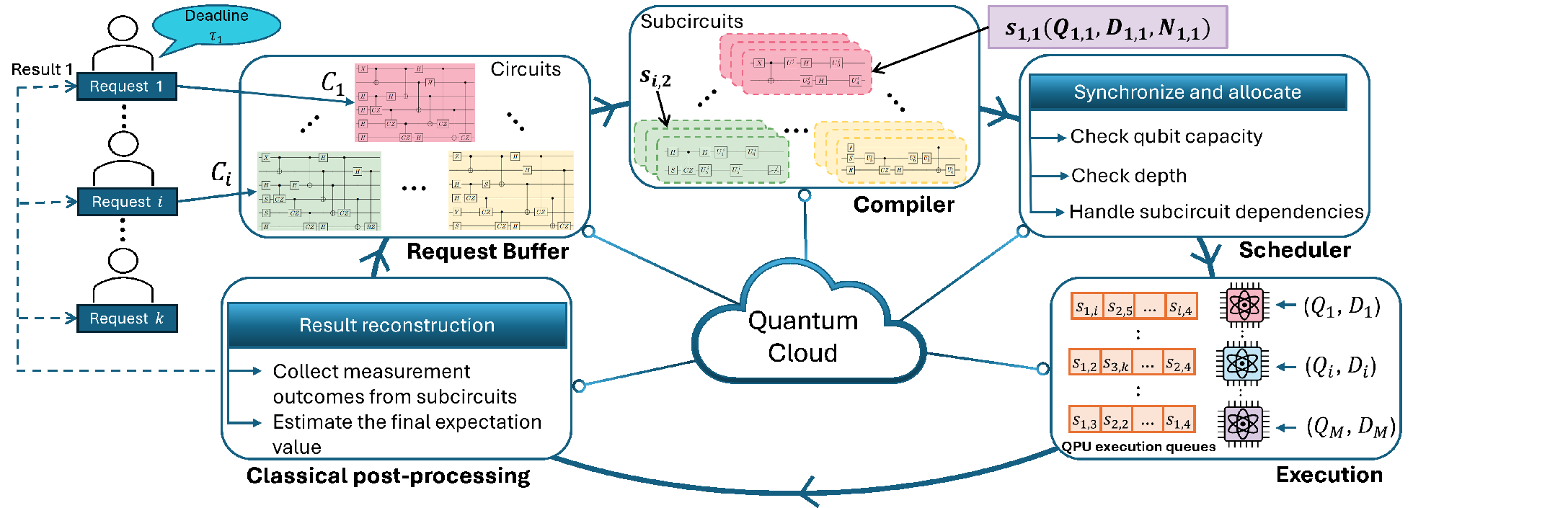}\vspace{-0.1in}
\caption{End-to-end near-term quantum cloud DQC workflow. User circuits with deadlines enter a request buffer, are cut into subcircuits by the compiler, scheduled on eligible QPUs under qubit count, depth and precedence constraints. Then, final results are reconstructed with classical post-processing.}
\label{fig_system_model}
\end{center}\vspace{-0.25in}
\end{figure*}

\section{System Model}\label{sec_system_model}
We consider a near-term quantum cloud, shown in Fig.~\ref{fig_system_model}, with a set $\mathcal{M}$ of M heterogeneous QPUs. Each QPU $ m \in \mathcal{M}$ is characterized by a qubit capacity $Q_m$ and a maximum feasible circuit depth $D_m$ that maintains acceptable fidelity. The QPUs are interconnected via classical communication links only, reflecting today's near-term capabilities in which quantum networks are not yet mature enough to provide reliable entanglement between QPUs \cite{BARRAL2025100747,LOCC_QPU}. 

The quantum cloud serves multiple users who submit a set $\mathcal{U}$ of U quantum computing requests. Each request $ i \in \mathcal{U}$ is for running a quantum circuit $C_i$ coming from a certain application (e.g., quantum optimization, search, order finding, sensing, ..., etc). Each circuit $C_i$ of request $i \in \mathcal{U}$ arrives with a user-specified deadline $\tau_{i}$ that specifies the maximum duration by which the circuit must be completed. The submitted circuits are collected by the request buffer in the quantum cloud where jobs are stored before compilation (Fig.~\ref{fig_system_model}) .

\vspace{-0.5cm}
\subsection{Quantum Cloud Compiler}\vspace{-0.2cm}
The cloud's compiler (Fig.~\ref{fig_system_model}) partitions each circuit $C_i$ into subcircuits  $\mathcal{S}_{i} = \{s_{i,1},...,s_{i,k}\}$, with $k = |\mathcal{S}_i|$, using wire cuts \cite{LOCC_wirecut} and gate cuts \cite{gate_cut}. This partitioning is performed only on large circuits that cannot be executed on a single QPU, which is the scenario this work focuses on. The compiler  ensures that the hardware constraints of the QPUs, such as qubit count and depth capacity, are met for at least one QPU for each $s_{i,j} \in S_i$ to make sure all subcircuits are feasible. Each subcircuit $s_{i,j}$ is described by its qubit demand $Q_{i,j}$, which is the number of qubits needed for the circuit, depth $D_{i,j}$, which is the number of its layers, and a shot count $N_{i,j}$, which is the number of shots, or samples, that need to be run in order to obtain an average result from the circuit. 

Specifically, when the compiler performs a gate cut, it replaces a two-qubit gate $V$ through quasiprobability decomposition into a weighted sum of local operations \cite{gate_cut}:
\begin{equation}
    V \;=\; \sum_{r=1}^{R}  a_r \,\big( A_r \otimes B_r \big), \qquad a_r \in \mathbb{R}
    \label{gatecut}
\end{equation}
where each term  $r \in \{1,\dots,R\}$ in the decomposition corresponds to a pair of local single-qubit gates applied on two separate partitions, resulting in a total of $2R$ subcircuits. These subcircuits are executed independently, and their outcomes are classically recombined at the last stage of circuit cutting workflow, as shown in Fig.~\ref{fig_system_model}, using the quasiprobability weights $\{a_r\}$ to estimate the expectation value of the target observable. Because quantum measurements are intrinsically probabilistic, observables are estimated from repeated, independent runs, called \say{shots}. However, gate cutting increases estimator variance, causing a sampling overhead where the required number of shots is multiplied by $(\sum_{r \in R} |a_r|)^2$ to reach the same accuracy as executing the nonlocal gate directly. 

Similarly, when the compiler performs a wire cut, it replaces a qubit wire with a linear combination of measure-prepare subcircuits \cite{Peng_2020}:
\begin{equation}
 \mathcal{I} \;=\; \sum_{b \in \Lambda} c_b \,\big( \mathcal{P}_b \circ \mathcal{M}_b \big),  \qquad c_b \in \mathbb{R}  
\end{equation}
where $\mathcal{I}$ is the single-qubit identity channel, $\mathcal{M}_b$ is a measurement operator in basis $b \in \Lambda$ with $\Lambda = \{I,X,Y,Z\}$, $\mathcal{P}_b$ is the preparation channel that initializes the quantum state in the same basis $b$, and $c_b$ are the quasiprobability decomposition weights. As with gate cuts, outcomes from the induced subcircuits are recombined using the quasiprobability weights $\{c_b\}$ and the shot overhead increases by $\bigl(\sum_{b}|c_b|\bigr)^{2}$. 


To reduce the number of subcircuits and sampling overhead in the compiling process, we adopt the recently developed shot-efficient LOCC wire cutting technique \cite{LOCC_wirecut}. Reducing the sampling overhead decreases the number of shots that must be executed to obtain a satisfying accuracy of the executed subcircuit. However, this induces a dependency in the execution since the measurement subcircuit must be executed before its corresponding preparation subcircuit. This is because the preparation subcircuit takes the classical measurement outcome as input. We represent the execution dependencies between the subcircuits by a precedence directed acyclic graph (DAG) $(\mathcal S_i,\, \mathcal P_i)$, where the vertices are the subcircuits $S_i$ of circuit $C_i$ and the edge set is $\mathcal{P}_i \subseteq \mathcal{S}_i \times \mathcal{S}_i$ defining the precedence relationships between subcircuits. We write $s_{i,u} \rightarrow s_{i,v} \in \mathcal{P}_i$ when $s_{i,v}$ must be executed after $s_{i,u}$. 


\subsection{Quantum Cloud Scheduler and Post-processing}
Thus, the compiler partitions the circuits in the request buffer using gate cuts and  LOCC wire cuts. As shown in Fig.~\ref{fig_system_model}, the cloud's scheduler then receives from the compiler, for each circuit $C_i$, the subcircuits $s_{i,j}$ with parameters $(Q_{i,j},\, D_{i,j},\, N_{i,j})$.
The scheduler estimates the per-shot runtime for each subcircuit as the total duration of subcircuit's layers, where each layer represents that set of gates which can be executed simultaneously \cite{Notads}. The time of a layer is the time of the longest gate in that level, with two-qubit gates being assumed to dominate single-qubit gates. We represent by $t_1$ the execution time of a single-qubit gate and by $t_2$ the execution time of a two-qubit gate, with $t2\gg t1$. Therefore, for a subcircuit $s_{i,j}$ with $\kappa_{1,i,j}$ single-qubit layers and $\kappa_{2,i,j}$ two-qubit layers the runtime is estimated as \cite{Notads}:
\begin{equation}
  T_{i,j} \;=\; \kappa_{1,i,j}\,t_1 \;+\; \kappa_{2,i,j}\,t_2, \qquad t_2 \gg t_1 .
\end{equation}
In addition, the scheduler receives from the compiler a precedence DAG $(\mathcal S_i,\, \mathcal P_i)$ capturing dependencies induced by LOCC wire cuts. Subject to this DAG, dependent subcircuits execute only after their predecessors (once the needed classical outcomes are available) whether sequentially on the same QPU or on a different QPU, while independent subcircuits may run in parallel across multiple QPUs. The scheduler assigns each subcircuit to a suitable QPU in terms of qubit count and depth and may distribute the shots of one subcircuit across multiple eligible QPUs. The precise optimzation model is presented in section~\ref{sec_optimization}.

Finally, each QPU executes its queue of assigned subcircuits and sends measurement outcomes to a classical workstation that recombines results using the quasiprobability weights: $c_b$ for wire cuts $a_r$ for gate cuts. Then, the classical workstation sends the final expectation values for each circuit to the corresponding user (Fig.~\ref{fig_system_model}).

 \section{Deadline-Aware Circuit Scheduling Optimization Formulation}\label{sec_optimization}

 In this section, we first derive the subcircuit and circuit execution time expressions. Then, we formulate a deadline-aware optimization framework that allows the distribution and optimization of subcircuit
shots across heterogeneous QPUs, while taking into account the LOCC wire-cut dependencies, QPU heterogeneity, to maximize the number of served requests of the different
users within their deadlines, while also minimizing the circuits completion time.

\subsection{Circuit Execution Time Calculation}

As discussed in Section ~\ref{sec_system_model}, we allow splitting the shots of a subcircuit across QPUs to increase parallelism and speed up execution. First, for each subcircuit $s_{i,j}$, we define the eligibility set $\mathcal{M}_{i,j} = \{m\in\mathcal{M}: Q_{i,j} \leq Q_m, D_{i,j} \leq D_m\}$ which has all the QPUs whose qubit and depth limits allow executing $s_{i,j}$. Then, we introduce a shot-allocation variable $y_{i,j,m}$  equal to the number of shots of $s_{i,j}$ executed on QPU $m \in \mathcal{M}_{i,j}$. We also introduce $z_i \in \{0,1\}$ which indicates whether circuit $C_i$ is served on time before its deadline ($z_i$ =1) or discarded ($z_i=0$).
We refer to the execution of subcircuit $s_{i,j}$ for $y_{i,j,m}$ shots on QPU $ m \in \mathcal{M}_{i,j}$  as the \emph{fragment} $(i,j,m)$. Each \emph{fragment} $(i,j,m)$ has a beginning execution time $b_{i,j,m}$, which is a control variable, and an end execution time defined using the compiler's per-shot runtime $T_{i,j}$ as
\begin{equation}\small
  e_{i,j,m}(b_{i,j,m},y_{i,j,m},z_i) \;=\;
\begin{cases}
0, & z_i=0,\\
b_{i,j,m} + T_{i,j}\cdot y_{i,j,m}, & z_i=1,
\end{cases}  
\end{equation}
The completion time of a subcircuit $s_{i,j} \in \mathcal{S}_i$ with all its shots is the latest \emph{fragment} end-time over all eligible QPUs, 
\begin{equation}\footnotesize
e_{i,j}(b_{i,j,m},y_{i,j,m},z_i) \;=\;
\begin{cases}
\; \;\;0, & z_i=0,\\
\max\limits_{m\in\mathcal{M}_{i,j}} e_{i,j,m}(b_{i,j,m},y_{i,j,m},1), & z_i=1,
\end{cases}
\end{equation}
and the circuit $C_i$ completion time $\forall i \in \mathcal{U}$ is the latest subcircuit end-time,
\begin{equation}\footnotesize
    T_i^{comp}(b_{i,j,m},y_{i,j,m},z_i) \;=\;
\begin{cases}
\;\;\;0, & z_i=0,\\
\max\limits_{j=1,\dots,k} e_{i,j}(b_{i,j,m},y_{i,j,m},1), & z_i=1.
\end{cases}  
\end{equation}

\subsection{Subcircuits Scheduling Optimization Problem}

Now, we formulate the scheduling problem to allocate subcircuits across heterogeneous QPUs and synchronize their execution to maximize the number of circuits that are completed before their deadlines and to encourage early completion by awarding a bonus when a circuit finishes within 80\% of its deadline.
To maximize the served requests $i \in \mathcal{U}$, where $\bigr | \mathcal{U} \bigr | = U$, the controllable optimization variables in our proposed optimization problem are: 1) $\mathbf{z} =  [z_1,\ldots,z_i,\ldots,z_U]$,
2) $\mathbf{B} = \bigl[\,b_{i,j,m}\,\bigr]_{\,i\in\mathcal{U},\ j=1,\ldots,k,\ m\in\mathcal{M}_{i,j}}$,
 3) $\mathbf{Y} = \bigl[\,y_{i,j,m}\,\bigr]_{\,i\in\mathcal{U},\ j=1,\ldots,k, m\in\mathcal{M}_{i,j}}$. Accordingly, the deadline-aware and shot-aware scheduling optimization problem can be formulated as follows:
\vspace{-0.05in}
\begin{subequations}\label{eq:O1}
\begin{align}
\mathcal{O}1: \quad \max_{\mathbf{z},\mathbf{B},\,\mathbf{Y}} & \sum_{i \in \mathcal{U}} z_i \ + \alpha \sum_{i \in \mathcal{U}} \mathds{1}(T_i^{comp}  \leq 0.8 \cdot\tau_{i}) \label{eq:objective}\\[2pt]
 {\text{s.t.}}\quad
 & \sum_{m \in \mathcal M_{i,j}} y_{i,j,m} \;=\; N_{i,j}\cdot z_i  \quad \forall i,j \label{eq:constraint1}\\
 & z_i = 1 \;\implies\; T_i^{comp} \leq \tau_{i} \quad \forall i \in \mathcal{U}\label{eq:constraint2}\\
 & b_{i,v,m} \ge e_{i,u}   \;\ \forall i,\ \forall (s_{i,u}\!\to s_{i,v}) \in \mathcal P_i,\ \forall m \label{eq:constraint3}\\
&\resizebox{2.2in}{\height}{$y_{i,j,m} \cdot y_{i',j',m} \ge 0 \implies
 b_{i,j,m} \ge e_{i',j',m} \vee b_{i',j',m} \ge e_{i,j,m}$} 
\label{eq:constraint4}\\
 & y_{i,j,m}\in\mathbb{N},\quad b_{i,j,m}\in\mathbb{N},\quad e_{i,j,m}\in\mathbb{N}. 
\label{eq:constraint5}
\end{align}
\end{subequations}





Constraint \eqref{eq:constraint1} ensures that for every subcircuit $s_{i,j}$, all of its shots $N_{i,j}$, are fully allocated across eligible QPUs when the circuit $C_i$ is served on time; otherwise, all subcircuits are dropped by not allocating shots for them $(y_{i,j,m} =0)$. Constraint \eqref{eq:constraint2} ensures that if circuit $C_i$ is considered served on time ($z_i$ =1), its completion time must not exceed its deadline $\tau_i$. This model tries to maximize how many are on-time by maximizing $\sum_{i \in \mathcal{U}}z_i$. 
Constraint \eqref{eq:constraint3} makes sure that the precedence conditions from wire cuts are met. A successor subcircuit $ s_{i,v}$ (preparation subcircuit) cannot start any of its \emph{fragments} on any QPU until its required predecessor $s_{i,u}$ (paired measurement subcircuit) has finished all its shots \eqref{eq:constraint1}. Constraint \eqref{eq:constraint4} guarantees a QPU runs on at most one \emph{fragment} at a time. Finally, constraint \eqref{eq:constraint5} ensures nonnegativity for allocations and time variables.

\subsection{Proposed Solution}
\begin{algorithm}[t!]\small
\caption{Simulated Annealing Algorithm for $\mathcal{O}1$}
\label{simulated_annealing_algorithm}
\begin{algorithmic}[1]
\State Initialize the current solution $\mathbf{z},\ \boldsymbol{B},\ \boldsymbol{Y}$ to a feasible solution in the search space
\State Set initial temperature $\tau_{\mathrm{sol}}=\tau_0$
\State Define $K$, the number of performed iterations for each temperature level
\State Initialize the optimal solution $\mathbf{z^*},\ \boldsymbol{B^*},\ \boldsymbol{Y^*}$  to the current solution
\While{$\tau_{\mathrm{sol}}>\tau_{\min}$}
    \For{$k=1$ to $K$}        
            \State Generate random neighbors  $\mathbf{z'},\ \boldsymbol{B'},\ \boldsymbol{Y'}$ via the following priorities: 
            \Statex \hspace{1.5em}(1) \textbf{Add circuit:} set $z'_i = 1$ for some $i \in \mathcal{U}$ 
\Statex \hspace{1.5em}(2) \textbf{Permute selected circuits:} permute the selected circuits for scheduling.
\Statex \hspace{1.5em}(3) \textbf{Local moves on $Y$ and $B$:} change shot splits $Y'$ and start times $B'$
             \If{ generated neighbors $\mathbf{z'},\ \boldsymbol{B'},\ \boldsymbol{Y'}$ satisfy $\mathcal{O}1$ constraints \eqref{eq:constraint1}, \eqref{eq:constraint2}, \eqref{eq:constraint3} \eqref{eq:constraint4} and \eqref{eq:constraint5}} 
            \State Calculate $\Delta O = O(\mathbf{z'}, \boldsymbol{B'}, \boldsymbol{Y'}) 
    - O(\mathbf{z}, \boldsymbol{B}, \boldsymbol{Y})$
                \If{$\Delta O \geq 0$ \textbf{or} $\exp(\Delta O/ \tau_{\mathrm{sol}}) > r\in\mathrm{Uniform}[0,1]$}
                    \State Update  $\mathbf{z},\ \boldsymbol{B},\ \boldsymbol{Y}$ to  $\mathbf{z'},\ \boldsymbol{B'},\ \boldsymbol{Y'}$, respectively
                \EndIf
            \EndIf
    \EndFor
    \If  {$O( \mathbf{z'},\ \boldsymbol{B'},\ \boldsymbol{Y'}) \geq  \ O( \mathbf{z^*},\ \boldsymbol{B^*},\ \boldsymbol{Y^*} )$}
    \State Update the best solution  $\mathbf{z^*},\ \boldsymbol{B^*},\ \boldsymbol{Y^*}$  to  $\mathbf{z'},\ \boldsymbol{B'},\ \boldsymbol{Y'}$ , respectively
    \EndIf
    \State Update $\tau_{\mathrm{sol}}$ according to adopted cooling schedule, e.g., exponential cooling, as $\tau_{\mathrm{sol}} = \tau_{\mathrm{sol}} \cdot \alpha_{\mathrm{sol}}$ 
\EndWhile
\State \textbf{return} the best solution found,  $\mathbf{z^*},\ \boldsymbol{B^*},\ \boldsymbol{Y^*}$.
\end{algorithmic}
\end{algorithm}
To solve problem $\mathcal{O}_1$, we propose a simulated annealing algorithm, a metaheuristic that searches for a near-optimal solution in a large space (Algorithm~\ref{simulated_annealing_algorithm}). We start from a feasible solution $(\mathbf z,\mathbf Y,\mathbf B)$ that satisfies all constraints of $\mathcal{O}_1$, set it as both the current and the best solution, choose a high initial temperature, and run a fixed number of iterations at each temperature level. In each iteration, we create a neighbor in the following priority: (i) \emph{add a circuit} by including an unscheduled $C_i$ (set $z_i=1$) and fixing $(\mathbf Y,\mathbf B)$ accordingly. If no circuit can be added, (ii) \emph{permute the selected set for scheduling} by swapping one selected circuit with another dropped one and repairing $(\mathbf Y,\mathbf B)$. If neither works, (iii) we try a local move on $(\mathbf Y,\mathbf B)$ that change execution order, QPU assignment, or shot splitting, such as: pushing a \emph{fragment} forward or backward on its QPU to reduce gaps, moving a subcircuit to another QPU, or splitting subcircuit’s shots across QPUs. Every candidate neighbor $(\mathbf z',\mathbf Y',\mathbf B')$ is checked for feasibility and evaluated by calculating the difference in the objective function. If it is better, the neighbor is accepted; otherwise, we may still accept it with a probability given by the Metropolis rule to help escape local minima. The temperature is reduced using an exponential cooling schedule until it reaches $\tau_{min}$, and the best solution is returned.



\section{Simulation Results and Analysis}\label{sec_simulations}
\begin{table}[t]
\centering
\caption{\textsc{Summary of Simulation Parameters}}
\label{tab:parameters}
\resizebox{\columnwidth}{!}{%
\begin{tabular}{llr}
\toprule
\textbf{Parameter} & \textbf{Description} & \textbf{Value} \\
\midrule
$M$ & Number of QPUs in the quantum cloud & 5\\
$N_0$ & Original number of shots & 10000 \cite{shotsNb}\\
$t_1$ & Single-qubit gate execution time & $t_1 = 1$ \cite{Notads} \\
$t_2$ & Two-qubit gate execution time & $t_2 = 10$ \cite{Notads} \\
$\gamma^2_{g}$ & Sampling overhead for gate cut & 9\cite{gate_cut}\\
$\gamma^2_{lw}$ & Sampling overhead for LOCC wire cut & 9\cite{LOCC_wirecut}\\
$\gamma^2_{ow}$ & Sampling overhead for old wire cut & 16\cite{Peng_2020}\\
\bottomrule
\end{tabular}%
}
\label{paraneters}
\end{table}

\begin{table}[t]
\centering
\footnotesize
\setlength{\tabcolsep}{4pt}
\caption{\textsc{Comparison of scheduling methods with the optimal solution.}}

\begin{tabular}{lcccc}
\toprule
Method & $T_1^{\text{comp}}$ &  $T_2^{\text{comp}}$& Served Requests\\
\midrule
Optimal (exhaustive) & 190 & 300 & 2  \\
Proposed \ref{simulated_annealing_algorithm} & 192 & 300 & 2  \\
Greedy   & 134 & \textemdash & 1 \\
List     & 140 & \textemdash & 1 \\
Random & 240 & 338 & 2  \\
Shot-agnostic & 230 & \textemdash & 1             \\
Dependency-agnostic & 240 & \textemdash & 1 \\ 
Dependency-and-shot-agnostic & 240 & \textemdash & 1 \\ 
 
\bottomrule
\end{tabular}
\label{exhaustive}\vspace{-0.5cm}
\end{table}
In this section, we conduct simulations to evaluate the efficiency and performance of our proposed deadline-aware scheduling framework. Unless stated otherwise, we adopt the following \emph{default setup} for our experiment. We consider $M=5$ heterogeneous QPUs. For each QPU $m\in \mathcal{M}$, the qubit capacity $Q_m$ and the maximum feasible circuit depth $D_m$ are sampled from a uniform distribution such that $
Q_m \sim \mathrm{Uniform}[10,20] \text{ qubits \cite{fitcut}},
D_m \sim \mathrm{Uniform}[10,20] \text{ layers},
 \forall m \in \mathcal{M}$.
A compiler partitions each circuit $C_i$  of request $i \in \mathcal{U}$ into subcircuits using either a gate cut or a wire cut. Following prior work \cite{LOCC_wirecut} and \cite{Gatevirtualization}, we treat the compiler’s circuit-cutting decomposition time as negligible compared to QPU execution time. Moreover, we also treat the final classical post-processing, where results are weighted and averaged, as negligible relative to the dominant QPU execution time, which is primarily affected by the sampling overhead and the number of shots.
We consider CNOT gate cutting, where a decomposed gate produces 12 independent subcircuits with no precedence \cite{gate_cut}. The LOCC wire cut produces 6 subcircuits with 3 measurement and 3 preparation subcircuits. A measurement subcircuit must finish before its corresponding preparation subcircuit, so we sample three precedence edges. We assume that classical communication between measurement and preparation subcircuits does not significantly increase the execution runtime, so can be practically ignored \cite{LOCC_wirecut}. In contrast, for the old wire cut method without dependencies, it results in 16 independent subcircuits \cite{Peng_2020,cutqc}.
For each subcircuit $s_{i,j} \in S_i$, we sample $Q_{i,j}$ and $D_{i,j}$ from uniform distributions, such that $ Q_{i,j} \sim \mathrm{Uniform}[5,20]\text{ qubits \cite{cutqc}, and}\, 
D_{i,j} \sim \mathrm{Uniform}[5,20]\text{ layers \cite{optimal_wire&gate}}
$, respectively, until the subcircuit is eligible on at least one QPU to ensure feasibility. Additionally, we sample each subcircuit’s single-qubit and two-qubit layer counts uniformly, where
$\kappa_{1,i,j} \sim  \mathrm{Uniform}[2,10]$ single-qubit gate layers, and $\kappa_{2,i,j} \sim \mathrm{Uniform}[3,10]$ two-qubit gate layers, respectively. The sampling is done while ensuring that $\kappa_{1,i,j} + \kappa_{2,i,j} = D_{i,j}$, and hence we set the subcircuit per-shot runtime as
\(
T_{i,j}=\kappa_{1,i,j}\,t_1+\kappa_{2,i,j}\,t_2,
\)  \cite{Notads}.
To obtain reliable expectation values with high precision, we fix an uncut baseline of $N_0$ shots per circuit, noting that achieving precision $\epsilon$ typically requires $N_0 = \mathcal{O}(1/\epsilon^{2})$~\cite{QEM}. 
After cutting, the required number of shots scales with the sampling overhead factor $\gamma^{2}$ of the chosen cut method, so the total shot budget per circuit becomes
$ N_0 \,\gamma^{2}$. 
We divide the circuit's total shots across the resulting subcircuits in proportion to the quasiprobability weights. For the cut methods considered in our work, these weights are equal, so the shots are split evenly across subcircuits \cite{LOCC_wirecut,Peng_2020,gate_cut}. The gates execution time, $N_0$, and sampling overhead values adopted in the default setup, are shown in Table \ref{tab:parameters}. 
Finally, for each circuit $C_i$, we compute a lower bound $T^{\mathrm{ref}}_i$ which is the smallest achievable completion time, assuming its subcircuits may execute simultaneously in parallel across eligible QPUs while respecting LOCC wire-cut precedency. Then, to model different user urgency levels, we sample a deadline coefficient
\(
d_{c,i} \sim \mathrm{Uniform}[3,\,10],
\)
and set the actual deadline as
\(
\tau_i \;=\; d_{c,i}\, T^{\mathrm{ref}}_i, \forall i \in \mathcal{U}.
\)

In order to evaluate the performance of our \emph{proposed} framework and solution algorithm, we compare it against several state-of-the-art scheduling algorithms. For a fair comparison, we consider all benchmark algorithms to be deadline-aware, although they are not originally designed to consider the deadlines associated with the incoming requests \cite{10821222,chandra2024network,ferrari2024design}. The benchmarks we compare against are as follows: 1) \emph{greedy:} earliest-deadline circuits are served first, and subcircuits can split their shots evenly across free eligible QPUs while respecting precedence constraints \cite{chandra2024network}, 2) \emph{random:} a scheduling algorithm that builds feasible schedules by randomly ordering subcircuits and assigning them to eligible QPUs, and can also stochastically split a subcircuit’s shots across multiple eligible QPUs, 3) \emph{list:} which follows the classic list scheduling algorithm, processing subcircuits in a precedence-respecting order and evenly splitting shots across free eligible QPUs \cite{ferrari2024design,10821222}, 4) \emph{shot-agnostic:} which represents the state-of-the-art approach in which shots are not distributed, treating each subcircuit’s shots as a single job, 5) \emph{dependency-agnostic:} which uses the widely adopted wire-cut method \cite{Peng_2020} instead of LOCC wire cuts, obtaining independent subcircuits, 6) \emph{dependency-and-shot-agnostic:} also uses the wire-cut method \cite{Peng_2020} and never splits a subcircuit’s shots between QPUs. We compare all these models with our \emph{proposed} scheduling framework which is both shot and dependency-aware.

We begin our simulations by running an exhaustive search solver to the \emph{proposed} optimization problem $\mathcal{O}_1$ to identify the \emph{optimal} schedule, and compare it against our \emph{proposed} simulated-annealing solver and the other benchmarks. We consider an instance with two requested circuits $(C_1,C_2)$ to be cut and scheduled with original $N_0=10$ shots per circuit. Table \ref{exhaustive} summarizes the results, where the optimal schedule served both circuits with $T_1^{\text{comp}}=190$ and  $T_2^{\text{comp}}=300$. Our \emph{proposed} framework also served both circuits, matching completion time of $C_2$ exactly and finishing $C_1$ at $T_1^{\text{comp}}=192$, which is a negligible $1.05\%$ performance gap. The \emph{random} scheduler also served both circuits but with significantly longer completion times. The other benchmarks served at most one circuit in this setup because $C_2$ misses its deadline. In terms of runtime, our \emph{proposed} approach produced a near-optimal schedule with a 97.9\% runtime reduction compared to the optimal exhaustive search solution. Accordingly, we use the \emph{proposed} scheduler for the remaining experiments at larger scales without running the exhaustive search solution.

\begin{figure}
\begin{tikzpicture}
  \begin{axis}[
    width=1\columnwidth, height=8.5cm,
    xlabel={Number of requests}, ylabel={Average served requests (\%)},
    ymin=0, ymax=80, ytick={0,20,40,60,80},
    xtick={1,2,3,4,5}, xticklabels={2,3,4,5,6},
    ymajorgrids=true, grid=major, grid style={dashed,gray!30},
    bar width=5pt, ybar, enlarge x limits=0.16,
    tick style={line width=0.8pt}, line width=1pt,
    legend columns=1,
    legend style={
      at={(0.7,1)}, anchor=north,
      draw=none, fill=white, fill opacity=0.9, font=\scriptsize,               
      row sep=1pt, column sep=4pt,     
      nodes={inner sep=1pt, outer sep=0pt, text depth=0.2ex},
    },
    legend image code/.code={
      \draw[#1,fill=#1,draw=black,line width=0.15pt] (0cm,-0.1cm) rectangle (0.20cm,0.10cm);
    },
    tick label style={font=\small}, label style={font=\small},
  ]


  \addplot+[ybar, bar shift=-15pt, fill=lgreen, draw=black, line width=0.15pt]
    coordinates {(1,72.3) (2,56.3) (3,50.0) (4,45.9) (5,43.0)};
  \addlegendentry{Proposed}

  \addplot+[ybar, bar shift=-10pt, fill=pink, draw=black, line width=0.15pt]
    coordinates {(1,67.5) (2,52.7) (3,48.3) (4,43.7) (5,40.7)};
  \addlegendentry{Shot-agnostic}
  \addplot+[ybar, bar shift=-5pt, fill=violet, draw=black, line width=0.15pt]
    coordinates {(1,61.4) (2,50.0) (3,42.9) (4,37.8) (5,34.6)};
  \addlegendentry{Dependency-agnostic}
  
  \addplot+[ybar, bar shift=0pt, fill=nviolet, draw=black, line width=0.15pt]
    coordinates {(1,58.1) (2,48.1) (3,41.8) (4,37.5) (5,34)};
  \addlegendentry{Dependency-and-shot-agnostic}

  \addplot+[ybar, bar shift=5pt, fill=lred, draw=black, line width=0.15pt]
    coordinates {(1,50.6) (2,35.1) (3,26.0) (4,20.2) (5,17.1)};
  \addlegendentry{Greedy}

  \addplot+[ybar, bar shift=10pt, fill=lblue, draw=black, line width=0.15pt]
    coordinates {(1,47.9) (2,34.2) (3,25.6) (4,20.5) (5,16.8)};
  \addlegendentry{List}

  \addplot+[ybar, bar shift=15pt, fill=lyellow, draw=black, line width=0.15pt]
    coordinates {(1,42.5) (2,33.3) (3,25.0) (4,21.8) (5,18.0)};
  \addlegendentry{Random}
  \end{axis}
\end{tikzpicture}\vspace{-0.2cm}
\caption{Average served requests vs. Number of requests.}
\label{fig:histogram_average_served_requests}\vspace{-0.6cm}
\end{figure}
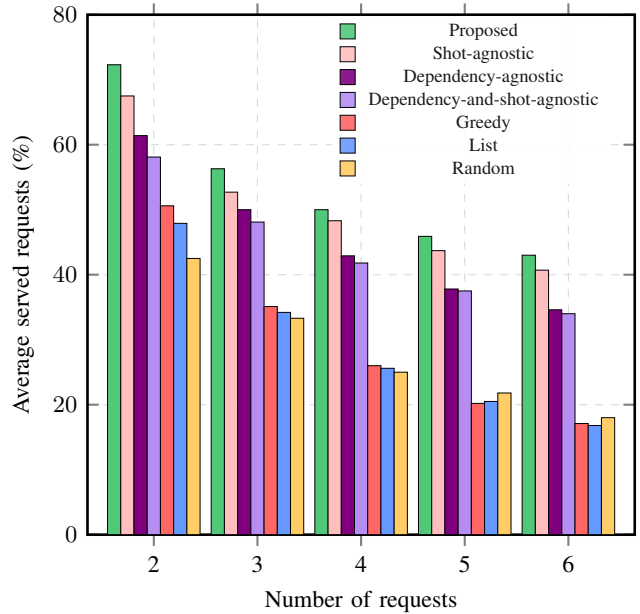

Next, in Fig.  \ref{fig:histogram_average_served_requests}, we examine the effects of increasing the number of requests on the average served requests for the different benchmarks over 50 independent Monte Carlo runs. We vary the number of user requests from 2 to 6 and observe from Fig. \ref{fig:histogram_average_served_requests} that the average served requests decreases as the number of requests grows. This trend is expected: with more requests, a larger number of subcircuits compete for the same limited QPU queue space, leading to higher contention. Across all evaluated number of requests, our \emph{proposed} approach consistently achieves the highest average number of served requests. Specifically, it outperforms the \emph{shot-agnostic} baseline by an average of 2.92\% (ranging from 1.7\% to 4.8\%), the \emph{dependency-agnostic} baseline by 8.16\% on average (with gains between 6.3\% and 10.9\%), and the \emph{dependency-and-shot-agnostic} baseline by 9.60\% on average (ranging from 8.2\% to 14.2\%). The improvements over the heuristic benchmarks are even more pronounced, reaching 23.7\% over \emph{greedy}, 24.5\% over \emph{list}, and 25.38\% over \emph{random} scheduling algorithms.

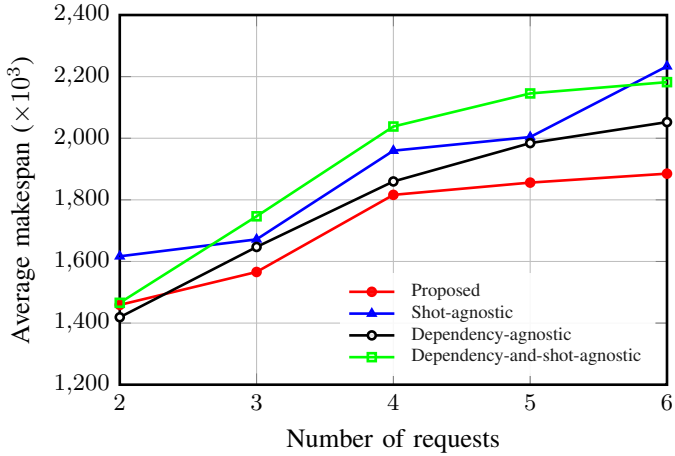
\begin{figure}[t!]  
\centering
\begin{tikzpicture}
    \begin{axis}[
      width=1\columnwidth,
      height=6.5cm,
      xlabel={Number of requests},
      ylabel={Average makespan ($\times 10^{3}$)},
      grid=major,
      tick label style={font=\small},
      line width=1pt,
      mark size=1.5pt,
      xmin=2, xmax=6,
      ymin=1200, ymax=2400,
      xtick={2,3,4,5,6},
      ytick={1200,1400,1600,1800,2000,2200,2400},
legend style={
  at={(0.98,0.02)},      
  anchor=south east,
  draw=none, fill=white, fill opacity=0.9,
  font=\scriptsize,
  cells={anchor=west},
  /tikz/column sep=4pt, /tikz/row sep=-2pt
},
legend columns=1,
    ]

      \addplot[red, solid, mark=*, mark options={solid, fill=red, draw=red}] 
        coordinates {(2,1459.020) (3,1565.912) (4,1816.209) (5,1855.991) (6,1885.057)};
      \addlegendentry{Proposed}

      \addplot[blue, solid, mark=triangle, mark options={solid, fill=white, draw=blue}] 
        coordinates {(2,1616.764) (3,1671.75) (4,1959.631) (5,2003.919) (6,2233.565)};

    \addlegendentry{Shot-agnostic}  
           
      \addplot[black, solid, mark=*, mark options={solid, fill=white, draw=black}] 
        coordinates {(2,1418.954) (3,1647.071) (4,1859.716) (5,1984.22) (6,2052.378)
};
      \addlegendentry{Dependency-agnostic}
 
          \addplot[green, solid, mark=square, mark options={solid, fill=green, draw=green}] 
        coordinates {(2,1465.801) (3,1746.614) (4,2038.041) (5,2145.436) (6,2182.107)};

    \addlegendentry{ Dependency-and-shot-agnostic}
    \end{axis}
\end{tikzpicture}\vspace{-0.2cm}
\caption{Average makespan vs number of requests.}
\label{fig:makespan}\vspace{-0.5cm}
\end{figure}

Next, Fig.~\ref{fig:makespan} reports the average makespan as the number of requests varies, using the same setup as Fig.~\ref{fig:histogram_average_served_requests}. As expected, makespan increases for all benchmarks as queues grow longer with higher load. Under light load (2 requests), the \emph{dependency-agnostic} scheme is faster than \emph{proposed} by $2.8\%$, since the wire-cutting method of~\cite{Peng_2020} produces independent subcircuits that can start immediately without LOCC precedence constraints. In contrast, the \emph{proposed} approach enforces LOCC dependencies between measurement and preparation subcircuits, requiring each prepare subcircuit to wait for its paired measurement subcircuit to finish. As the number of requests increases, however, the \emph{proposed} framework achieves a lower makespan despite respecting these dependencies. Its LOCC wire cut generates fewer subcircuits and imposes a smaller sampling overhead, reducing the total amount of work queued on the QPUs. Moreover, the waiting periods of preparation subcircuits become naturally masked at higher load, as subcircuits from other circuits occupy the QPUs during those intervals. On average, \emph{proposed} reduces makespan by $4.34\%$ relative to \emph{dependency-agnostic} (ranging from $2.8\%$ at 2 requests to $8.2\%$ at 6 requests) frameworks. Compared to \emph{shot-agnostic}, it lowers makespan by an average of $9.28\%$ (range: $6.33\% \text{ to } 15.60\%$). Also, against \emph{dependency-and-shot-agnostic}, reductions vary from $0.46\%$ to $13.61\%$, averaging $9.76\%$. Taken together, Figs.~\ref{fig:histogram_average_served_requests} and \ref{fig:makespan} show that the \emph{proposed} scheduler provides a clear advantage, since it generally serves more requests while achieving a shorter makespan than all competing baselines. The combination of efficient LOCC wire cuts and informed shot distribution keeps the QPUs more effectively utilized, translating into both higher served throughput and faster overall completion times.

\begin{figure}[!t]
  \centering
  \begin{subfigure}[t]{0.9\columnwidth}
    \centering
    \includegraphics[width=\linewidth]{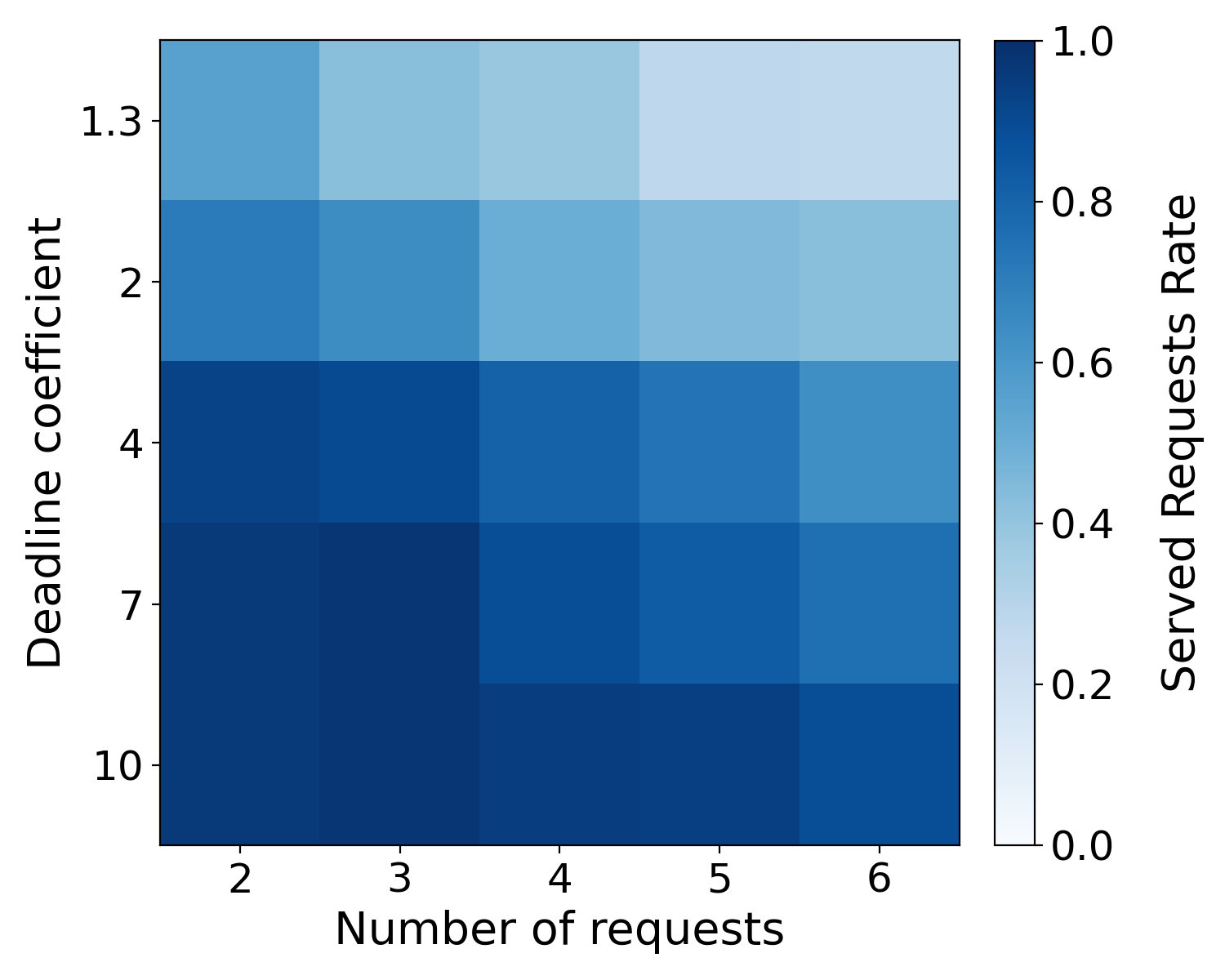}
    \caption{Proposed}
    \label{fig:heatmap_split}
  \end{subfigure}\hfill
  \begin{subfigure}[t]{0.9\columnwidth}
    \centering
    \includegraphics[width=\linewidth]{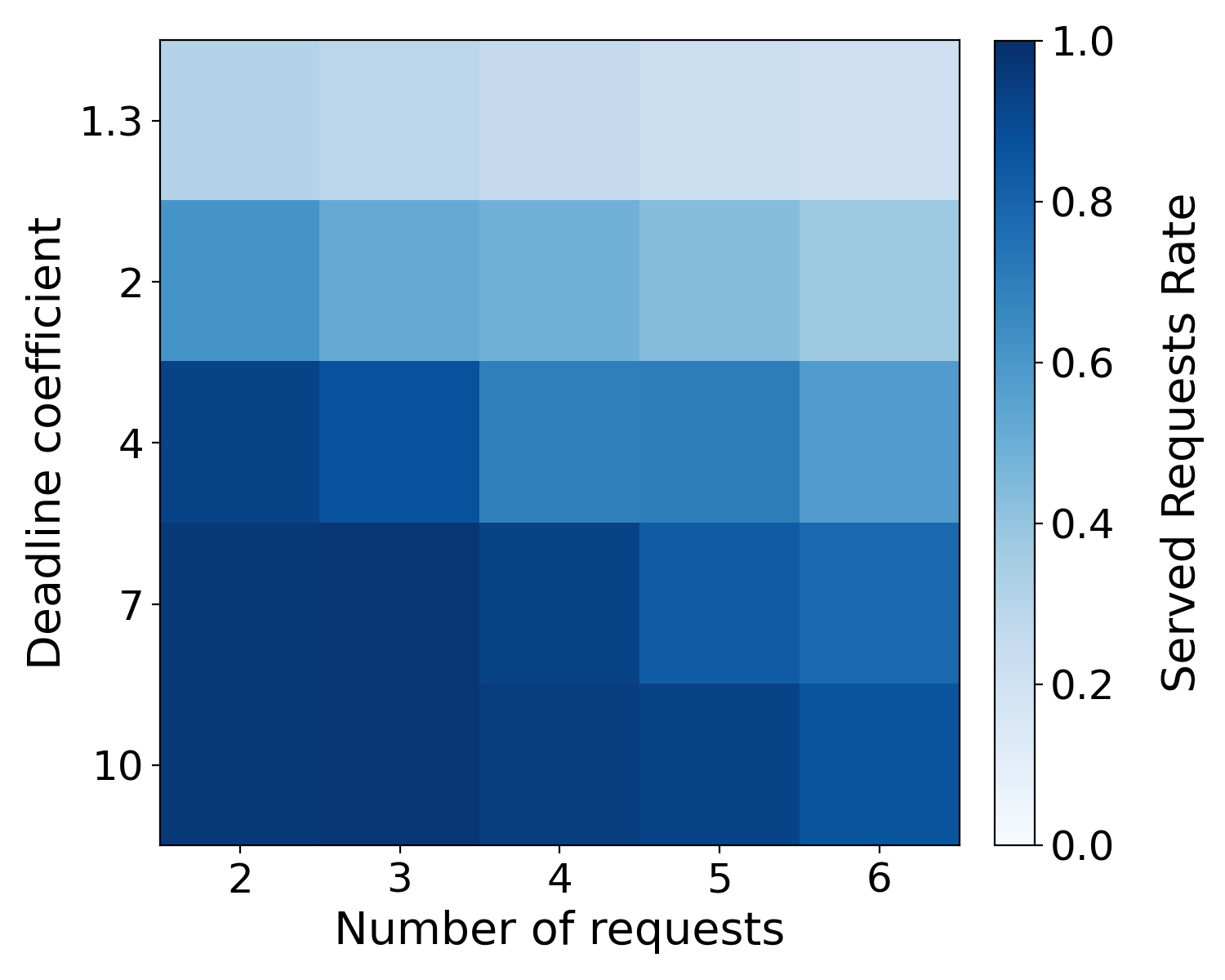}
    \caption{Shot-agnostic}
    \label{fig:heatmap_nosplit}
  \end{subfigure}
  \caption{Served requests rate vs deadline factor and number of requests.}
  \label{fig:heatmap_both}
  \vspace{-0.35in}
\end{figure}

Back to Fig.~\ref{fig:histogram_average_served_requests}, we observe that the performance gap between the \emph{proposed} and \emph{shot-agnostic} frameworks is relatively modest, with an average difference of $2.92\%$ in served requests. This behavior is primarily due to the randomized deadline coefficients, which were sampled from the interval 
$[3,10]$. The presence of multiple circuits with generous deadlines reduces the urgency of scheduling decisions, allowing the \emph{shot-agnostic} scheme to admit only slightly fewer requests than the \emph{proposed} method under these mixed-deadline conditions. To more systematically investigate the influence of deadline tightness, we, next, conduct an experiment that jointly varies the deadline coefficient and the number of user requests, and show the results in Fig.~\ref{fig:heatmap_both}. Using the same default experimental setup, we sweep the deadline factor $d_c \in \{1.3,\,2,\,4,\,7,\,10\}$, covering the full spectrum from very tight to very lenient deadlines. For each value of $d_c$, we vary the number of requests from 2 to 6. In each sweep, all requests share the same deadline factor, and we report the average number of served requests across multiple runs to capture the combined effect of system load and deadline urgency. Figs.~\ref{fig:heatmap_split} and \ref{fig:heatmap_nosplit} show that for both the \emph{proposed} and \emph{shot-agnostic} frameworks, the average number of served requests decreases as deadlines tighten. When deadlines are generous, the two approaches perform similarly: at $d_c = 10$, the \emph{proposed} method offers only a $0.8\%$ improvement, and at $d_c = 7$ the gain is $1.2\%$ on average. The differences become more pronounced as deadlines grow tighter. At $d_c = 4$, the \emph{proposed} approach outperforms \emph{shot-agnostic} by $4.8\%$ on average (and by $5\%$ at 6 requests). For $d_c = 2$, the average improvement increases to $6.2\%$, with a notable $13\%$ gain at 3 requests. Under the tightest constraint, $d_c = 1.3$, the advantage becomes substantial, reaching $12.8\%$ on average, including $14\%$ gains at 3 and 4 requests and a peak of $25\%$ at 2 requests. This demonstrates that the benefits of the \emph{proposed} scheme become increasingly significant as user deadlines grow more urgent.

\begin{figure}[!t]
  \centering
  \begin{subfigure}[t]{0.95\columnwidth}
    \centering
    \includegraphics[width=\linewidth]{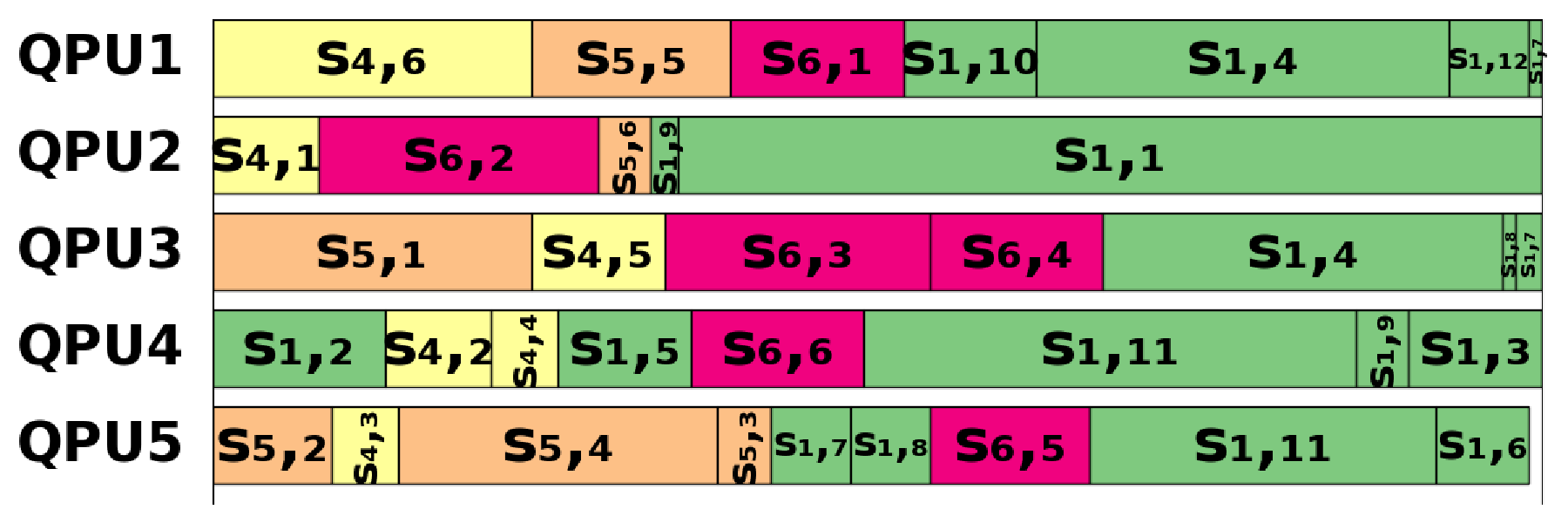}
    \caption{Proposed}
    \label{fig:gant1}
  \end{subfigure}\hfill
  \begin{subfigure}[t]{0.95\columnwidth}
    \centering
    \includegraphics[width=\linewidth]{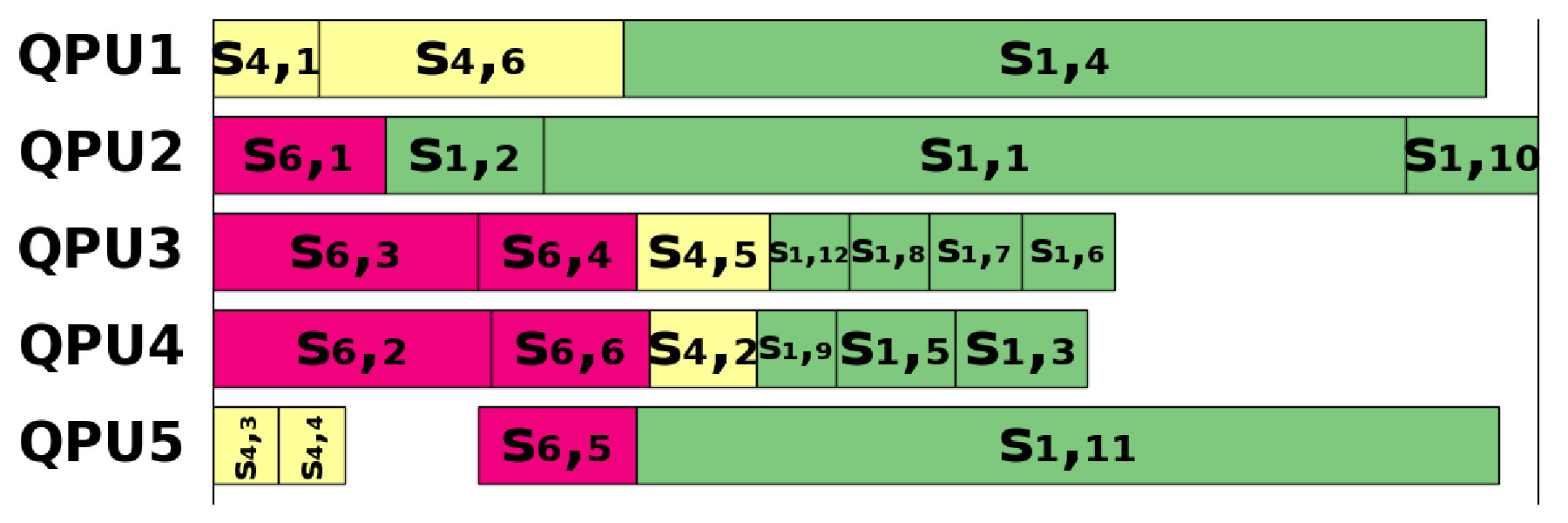}
    \caption{Shot-agnostic }
    \label{fig:gant2}
  \end{subfigure}
    \begin{subfigure}[t]{0.95\columnwidth}
    \centering
    \includegraphics[width=\linewidth]{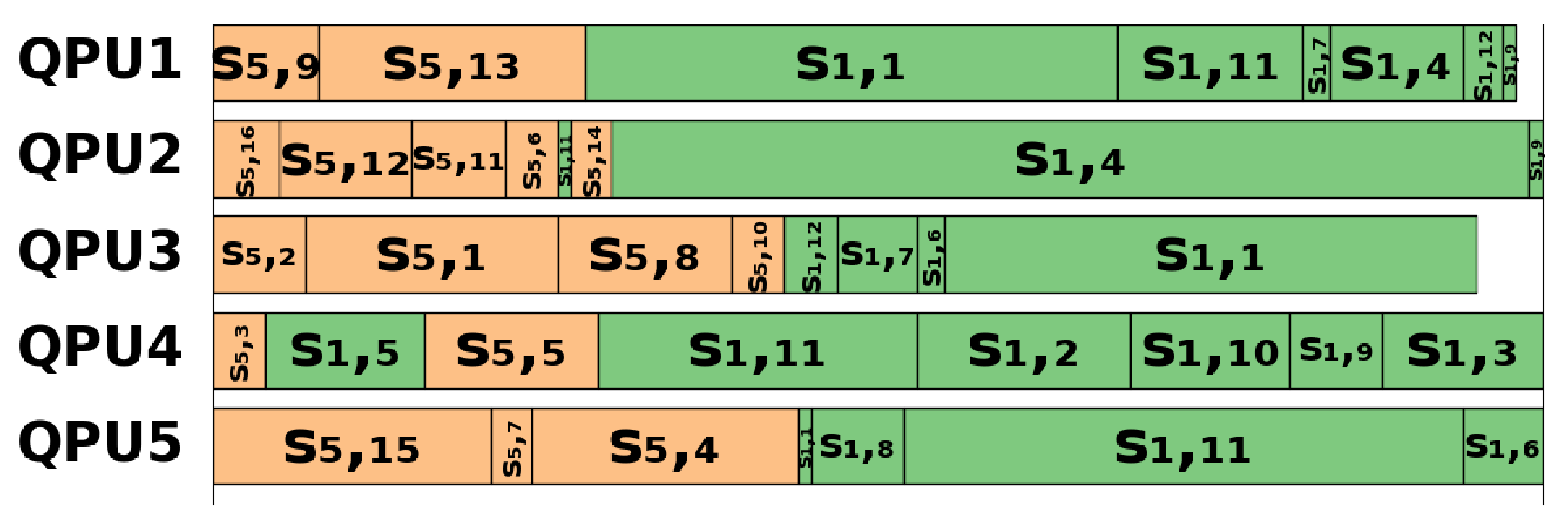}
    \caption{Dependency-agnostic.}  
    \label{fig:gant3}
  \end{subfigure}\hfill
    \begin{subfigure}[t]{0.95\columnwidth}
    \centering
    \includegraphics[width=\linewidth]{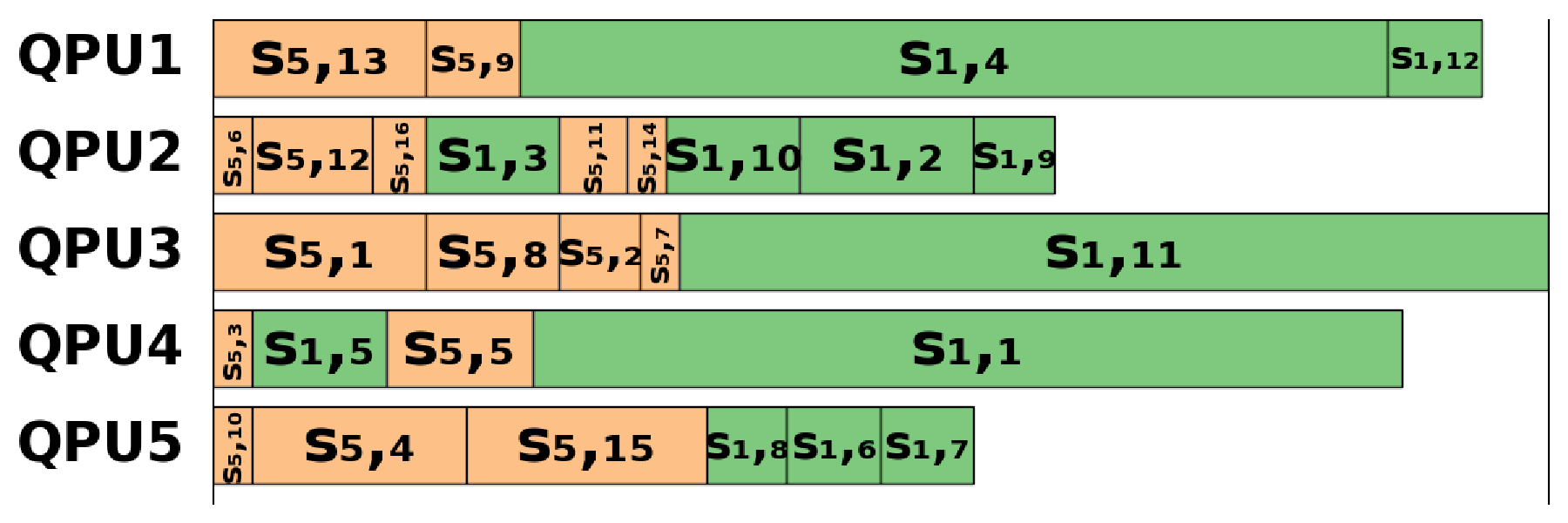}
    \caption{Dependency-and-shot-agnostic.}
    \label{fig:gant4}
  \end{subfigure}
  \vspace{-0.05in}
  \caption{Scheduling outputs of different benchmarks.}
  \label{fig:gant}\vspace{-0.5cm}
\end{figure}

Finally, in Fig.~\ref{fig:gant} we visualize the schedules produced by different benchmarks for a single instance with six circuits in the request buffer, where $C_1$ and $C_2$ are partitioned via gate cuts, while $C_3$–$C_6$ use wire cuts. As shown in Figs.~\ref{fig:gant1} and \ref{fig:gant2}, the circuits partitioned via LOCC wire cuts yield six subcircuits each, with precedence respected so that a prepare fragment starts only after its paired measurement subcircuit finishes. Specifically, for $C_4$ the edges are $(s_{4,1}\!\to\!s_{4,2})$, $(s_{4,3}\!\to\!s_{4,4})$, and $(s_{4,6}\!\to\!s_{4,5})$; for $C_5$: $(s_{5,1}\!\to\!s_{5,6})$, $(s_{5,2}\!\to\!s_{5,5})$, and $(s_{5,4}\!\to\!s_{5,3})$; and for $C_6$: $(s_{6,1}\!\to\!s_{6,4})$, $(s_{6,2}\!\to\!s_{6,6})$, and $(s_{6,3}\!\to\!s_{6,5})$. By contrast, the gate cut produces $12$ independent subcircuits with no precedence constraints. In this sample, \emph{proposed} admits four circuits, whereas \emph{shot-agnostic} serves only three. The tighter scheduling enabled by subcircuit shot splitting (e.g., $s_{1,4}$ split across QPU$_1$ and QPU$_3$) keeps the QPUs busier, providing enough capacity to admit a fourth circuit. In Figs.~\ref{fig:gant3} and \ref{fig:gant4}, both benchmarks admit only two circuits. Here, the precedence edges mentioned above are no longer a constraint, since the older wire-cut technique produces fully independent subcircuits. However, the wire cut increases the number of subcircuits from $6$ to $16$ in $C_5$ and increases the sampling overhead, causing fewer deadlines to be met.

\vspace{-0.2cm}
\section{Conclusion}\label{sec_conclusion}\vspace{-0.2cm}
In this paper, we have studied the scheduling of DQC circuits in near-term quantum clouds with heterogeneous QPUs under LOCC constraints. Specifically, we proposed a deadline-aware scheduling optimization framework for that optimizes the per-subcircuit shot distribution across QPUs, captures subcircuit dependencies, and minimizes runtime to serve as many circuits as possible. Furthermore, we conducted several experiments to study the impact of shot distribution and the effect of adopting LOCC wire cuts, finding that both contribute to more served requests and lower makespans. Going forward, we aim to jointly optimize circuit cutting and scheduling on heterogeneous QPUs.

\vspace{-0.35cm}
\begin{spacing}{0.88}
\bibliographystyle{IEEEtran}
\bibliography{References}

\begin{thebibliography}{10}
\providecommand{\url}[1]{#1}
\csname url@samestyle\endcsname
\providecommand{\newblock}{\relax}
\providecommand{\bibinfo}[2]{#2}
\providecommand{\BIBentrySTDinterwordspacing}{\spaceskip=0pt\relax}
\providecommand{\BIBentryALTinterwordstretchfactor}{4}
\providecommand{\BIBentryALTinterwordspacing}{\spaceskip=\fontdimen2\font plus
\BIBentryALTinterwordstretchfactor\fontdimen3\font minus \fontdimen4\font\relax}
\providecommand{\BIBforeignlanguage}[2]{{%
\expandafter\ifx\csname l@#1\endcsname\relax
\typeout{** WARNING: IEEEtran.bst: No hyphenation pattern has been}%
\typeout{** loaded for the language `#1'. Using the pattern for}%
\typeout{** the default language instead.}%
\else
\language=\csname l@#1\endcsname
\fi
#2}}
\providecommand{\BIBdecl}{\relax}
\BIBdecl

\bibitem{BARRAL2025100747}
D.~Barral \emph{et~al.}, ``Review of distributed quantum computing: From single qpu to high performance quantum computing,'' \emph{Computer Science Review}, vol.~57, p. 100747, 2025.

\bibitem{caleffi2024distributed}
M.~Caleffi, M.~Amoretti, D.~Ferrari, J.~Illiano, A.~Manzalini, and A.~S. Cacciapuoti, ``Distributed quantum computing: a survey,'' \emph{Computer Networks}, vol. 254, p. 110672, 2024.

\bibitem{chehimi2022physics}
M.~Chehimi and W.~Saad, ``Physics-informed quantum communication networks: A vision toward the quantum internet,'' \emph{IEEE network}, vol.~36, no.~5, pp. 32--38, 2022.

\bibitem{Peng_2020}
T.~Peng \emph{et~al.}, ``Simulating large quantum circuits on a small quantum computer,'' \emph{Phys. Rev. Let.}, vol. 125, no.~15, Oct. 2020.

\bibitem{gate_cut}
K.~Mitarai and K.~Fujii, ``Constructing a virtual two-qubit gate by sampling single-qubit operations,'' \emph{New Journal of Physics}, vol.~23, no.~2, p. 023021, Feb. 2021.

\bibitem{HPC}
M.~Tejedor, B.~Casas, J.~Conejero, A.~Cervera-Lierta, and R.~M. Badia, ``Distributed quantum circuit cutting for hybrid quantum-classical high-performance computing,'' \emph{arXiv preprint arXiv:2505.01184}, 2025.

\bibitem{LOCC_wirecut}
H.~Harada \emph{et~al.}, ``Doubly optimal parallel wire cutting without ancilla qubits,'' \emph{PRX Quantum}, vol.~5, no.~4, p. 040308, 2024.

\bibitem{10821222}
D.~Ferrari, M.~Bandini, and M.~Amoretti, ``Execution management of distributed quantum computing jobs,'' in \emph{2024 IEEE International Conference on Quantum Computing and Engineering (QCE)}, vol.~02, 2024, pp. 150--154.

\bibitem{ferrari2024design}
D.~Ferrari and M.~Amoretti, ``A design framework for the simulation of distributed quantum computing,'' in \emph{Proc. of the 2024 Workshop on High Performance and Quantum Computing Integration}, 2024, pp. 4--10.

\bibitem{bahrani2024resource}
S.~Bahrani \emph{et~al.}, ``Resource management and circuit scheduling for distributed quantum computing interconnect networks,'' \emph{arXiv preprint arXiv:2409.12675}, 2024.

\bibitem{chandra2024network}
N.~K. Chandra, E.~Kaur, and K.~P. Seshadreesan, ``Network operations scheduling for distributed quantum computing,'' in \emph{2024 IEEE 6th International Conference on Trust, Privacy and Security in Intelligent Systems, and Applications (TPS-ISA)}.\hskip 1em plus 0.5em minus 0.4em\relax IEEE, 2024, pp. 506--515.

\bibitem{EC2S}
Z.~Du, W.~Zhang, W.~Wei, J.~Chen, T.~Han, Z.~Liang, and Y.~Mao, ``Efficient circuit cutting and scheduling in a multi-node quantum system with dynamic epr pairs,'' \emph{arXiv preprint arXiv:2412.18709}, 2024.

\bibitem{larqucut}
L.~Liu, ``Larqucut: A new cutting and mapping approach for large-sized quantum circuits in distributed quantum computing (dqc) environments,'' \emph{ACM Transactions on Architecture and Code Optimization}, 2025.

\bibitem{optimal_wire&gate}
S.~Brandhofer, I.~Polian, and K.~Krsulich, ``Optimal partitioning of quantum circuits using gate cuts and wire cuts,'' \emph{IEEE Transactions on Quantum Engineering}, vol.~5, pp. 1--10, 2024.

\bibitem{luo2025adaptive}
W.~Luo, J.~Zhao, T.~Zhan, and Q.~Guan, ``Adaptive job scheduling in quantum clouds using reinforcement learning,'' \emph{arXiv preprint arXiv:2506.10889}, 2025.

\bibitem{Notads}
D.~Bhoumik, R.~Majumdar, A.~Saha, and S.~Sur-Kolay, ``Distributed scheduling of quantum circuits with noise and time optimization,'' \emph{arXiv preprint arXiv:2309.06005}, 2023.

\bibitem{cutandshoot}
G.~Bisicchia, A.~Bocci, J.~Garc{\'\i}a-Alonso, J.~M. Murillo, and A.~Brogi, ``Cut\&shoot: Cutting \& distributing quantum circuits across multiple nisq computers,'' in \emph{2024 IEEE International Conference on Quantum Computing and Engineering (QCE)}, vol.~2.\hskip 1em plus 0.5em minus 0.4em\relax IEEE, 2024, pp. 187--192.

\bibitem{LOCC_QPU}
A.~Carrera~Vazquez, C.~Tornow, D.~Riste, S.~Woerner, M.~Takita, and D.~J. Egger, ``Combining quantum processors with real-time classical communication,'' \emph{Nature}, vol. 636, no. 8041, pp. 75--79, 2024.

\bibitem{shotsNb}
T.-H. Chen and W.-H. Hung, ``Feasibility and limitations of generalized grover search algorithm-based quantum asymmetric cryptography: An implementation study on quantum hardware,'' \emph{Electronics}, vol.~14, no.~19, 2025.

\bibitem{fitcut}
S.~Kan \emph{et~al.}, ``Scalable circuit cutting and scheduling in a resource-constrained and distributed quantum system,'' in \emph{2024 IEEE International Conference on Quantum Computing and Engineering (QCE)}, vol.~1.\hskip 1em plus 0.5em minus 0.4em\relax IEEE, 2024, pp. 1077--1088.

\bibitem{Gatevirtualization}
N.~Tornow, E.~Giortamis, and P.~Bhatotia, ``Scaling quantum computations via gate virtualization,'' \emph{arXiv preprint arXiv:2406.18410}, 2024.

\bibitem{cutqc}
W.~Tang \emph{et~al.}, ``{Cutqc}: using small quantum computers for large quantum circuit evaluations,'' in \emph{Proceedings of the 26th ACM International conference on architectural support for programming languages and operating systems}, 2021, pp. 473--486.

\bibitem{QEM}
Z.~Cai \emph{et~al.}, ``Quantum error mitigation,'' \emph{Rev. Mod. Phys.}, vol.~95, p. 045005, Dec 2023.

\end{thebibliography}
\end{spacing}

\end{document}